\documentclass[journal,10pt,a4paper,onecolumn]{IEEEtran}
\usepackage{amsmath,amssymb,amsbsy,amsxtra,graphicx}
\usepackage{array,multirow}
\usepackage{threeparttable}
\usepackage{booktabs}
\usepackage{url}
\usepackage{cite}
\usepackage{psfrag}
\usepackage{adjustbox}
\usepackage{cite}
\usepackage{bm}
\usepackage{subcaption}
\usepackage{epstopdf}

\usepackage{algpseudocode}
\usepackage{algorithm}

\usepackage{textpos}

\def\A{{\mathbf A}}
\def\B{{\mathbf B}}
\def\I{{\mathbf I}}

\def\S{{\mathbf S}}

\def\0{{\mathbf 0}}

\def\x{{\mathbf x}}

\def\s{{\mathbf s}}

\def\b{{\mathbf b}}

\title{On Sparse Graph Fourier Transform}
\author{Seyed Hamid Safavi$^{\dagger \star}$, Manas Khatua$^{\ddagger}$, Ngai-Man Cheung$^{\dagger}$, Farah Torkamani-Azar $^{\star}$\\
$^{\dagger}$ Singapore University of Technology and Design, Singapore 
         $^{\star}$ Shahid Beheshti University, Iran, \\ 
         $^{\ddagger}$ Indian Institute of Technology, Jodhpur, India\\
         E-mail:\{ h\_safavi@sbu.ac.ir, manaskhatua@gmail.com, ngaiman\_cheung@sutd.edu.sg, f-torkamani@sbu.ac.ir\}
\thanks{ 
}}

\begin{document}
\maketitle

\begin{abstract}
In this paper,
we propose a new regression-based algorithm to compute 
Graph Fourier Transform (GFT).
Our algorithm allows different regularizations to be included
when computing the GFT analysis components, so that the resulting components can be tuned for a specific task. We propose using the lasso penalty in our proposed framework to obtain analysis components with sparse loadings.
We show that the components from this proposed {\em sparse GFT} can  identify and select correlated signal sources into sub-graphs, and perform frequency analysis {\em locally} within these sub-graphs of correlated sources.
Using real network traffic datasets,
we demonstrate that sparse GFT can achieve outstanding performance in an 
anomaly detection task.

\end{abstract} 

\begin{textblock*}{3cm}(8.5cm,-8cm)
	\makebox[0.1\columnwidth]{Presented at 3'rd Graph Signal Processing Workshop - GSP 18}
\end{textblock*}
\section{Introduction}
\label{sec:intro}

Our work addresses improvement in graph Fourier transform (GFT), an algorithm that is essential for analysis/processing of graph signals. Existing  work generally uses
eigendecomposition on the graph Laplacian matrix
to compute the analysis components of GFT~\cite{shuman2013emerging,ortega2017graph}.  The limitation is  that this does not allow adaptation of the components for different tasks and has limited flexibility. 

In this paper, we propose a  new regression-based algorithm for GFT. We
research using our algorithm to obtain GFT analysis components with sparse loading.
As will be discussed, the proposed sparse GFT has  a variable selection property
in addition to the 
oscillatory property as in ordinary GFT.
It selects correlated variables (signal sources) and provides graph frequency analysis within subgraphs of selected variables.


A few  improvements on GFT have been proposed recently, but their motivations and approaches are significantly different. In particular, they focus on fast GFT computation~\cite{le2017approximate}, 
GFT for directed graph~\cite{Sardellitti:2017,Shafipour:2017},
the irregular nature of the graph~\cite{girault:2018}, 
and 
dynamic graph signal with temporally-varying Laplacian matrices~\cite{villafane2017dynamic}.


\vspace{-0.3cm}
\section{Method}
\label{sec:method}
\vspace{-0.3cm}
\subsection{GFT}

We model the $p$-dimensional graph signal $\x$ as a signal residing on a graph with $p$ vertices. We focus on undirected weighted graphs, $G=\{V, E, W\}$, where $V$ is the set of $p$ vertices, $E$ is the set of edges and $W$ is the $p$-by-$p$ adjacency matrix of the graph.
$D$ is the $p$-by-$p$ degree matrix and is diagonal.
  The $p$-by-$p$ 
normalized graph Laplacian matrix $\Phi$ is defined to be  $I - D^{-1/2} W D^{-1/2}$.

The eigenvectors of the graph Laplacian have different oscillatory properties with respect to the graph~\cite{shuman2013emerging}. 
They are used in GFT for analysis of a graph signal and extraction of graph frequency components. Specifically, the analysis step of GFT is the projection of the graph signal $\x$ onto the analysis component $\b_m$:
$\tilde{\x}[m]  = \langle \x, \b_m \rangle$; 
$\tilde{\x}[m]$ is the $m$-th element of the vector of graph frequencies. Existing work generally uses eigendecomposition on the Laplacian matrix for computing the analysis component $\b_m$~\cite{shuman2013emerging}.  In what follows, we describe a new GFT algorithm which has much flexibility to incorporate different properties into $\b_m$, notably sparsity.

\vspace{-0.3cm}
\subsection{Algorithm for sparse GFT}

We propose a regression-based algorithm to compute GFT analysis component $\b_m$.
This regression-based algorithm allows us to include lasso penalty in the objective function so that we can obtain GFT analysis components with sparse loadings. 
Specifically, we apply the classical result from~\cite{Zou06}, which was used in~\cite{Zou06} to transform a general PCA problem into a regression-type problem.
{\bf Algorithm \ref{alg:regression}} lists our  proposed  regression-based algorithm for GFT.




\begin{algorithm}
\caption{Regression-based algorithm for GFT:} \label{alg:regression}
\begin{algorithmic}[1]
\State \textbf{Input}: $G=\{V, E, W\}$.
\State Compute the Laplacian matrix $\Phi$ from  $G$.
\State Perform factorization of the graph Laplacian $\Phi = \S^T \S$, to obtain a $h\times p$ matrix $\S$.
\State Perform regression on the matrix factor $\S$ using Theorem 3 of~\cite{Zou06}.
Let $\s_i$ be the row vector of $\S$, 
$\A$ and $\B$ be $p$-by-$k$ matrices, $\b_m$  are the columns of $\B$, and $\lambda > 0$.
We solve:
\vspace{-0.2cm}
\begin{equation}
\begin{aligned}
& \underset{\A,\B}{\text{minimize}}
& & \sum_{i=1}^h || \s_i - \A \B^T \s_i   ||^2  + \lambda \sum_{m=1}^k || \b_m ||^2   \\
& \text{subject to}
& & \A^T \A = \I_{k \times k}, \S^T \S=  \Phi
\end{aligned}
\label{eq:regression1}
\end{equation}
\State \textbf{Output:} $\b_m$, the columns of $\B$.
\end{algorithmic}
\end{algorithm}

\begin{figure*}[t]
\begin{minipage}[t]{0.49\linewidth}
  \centering
  \centerline{\includegraphics[width=8cm,height=7cm]{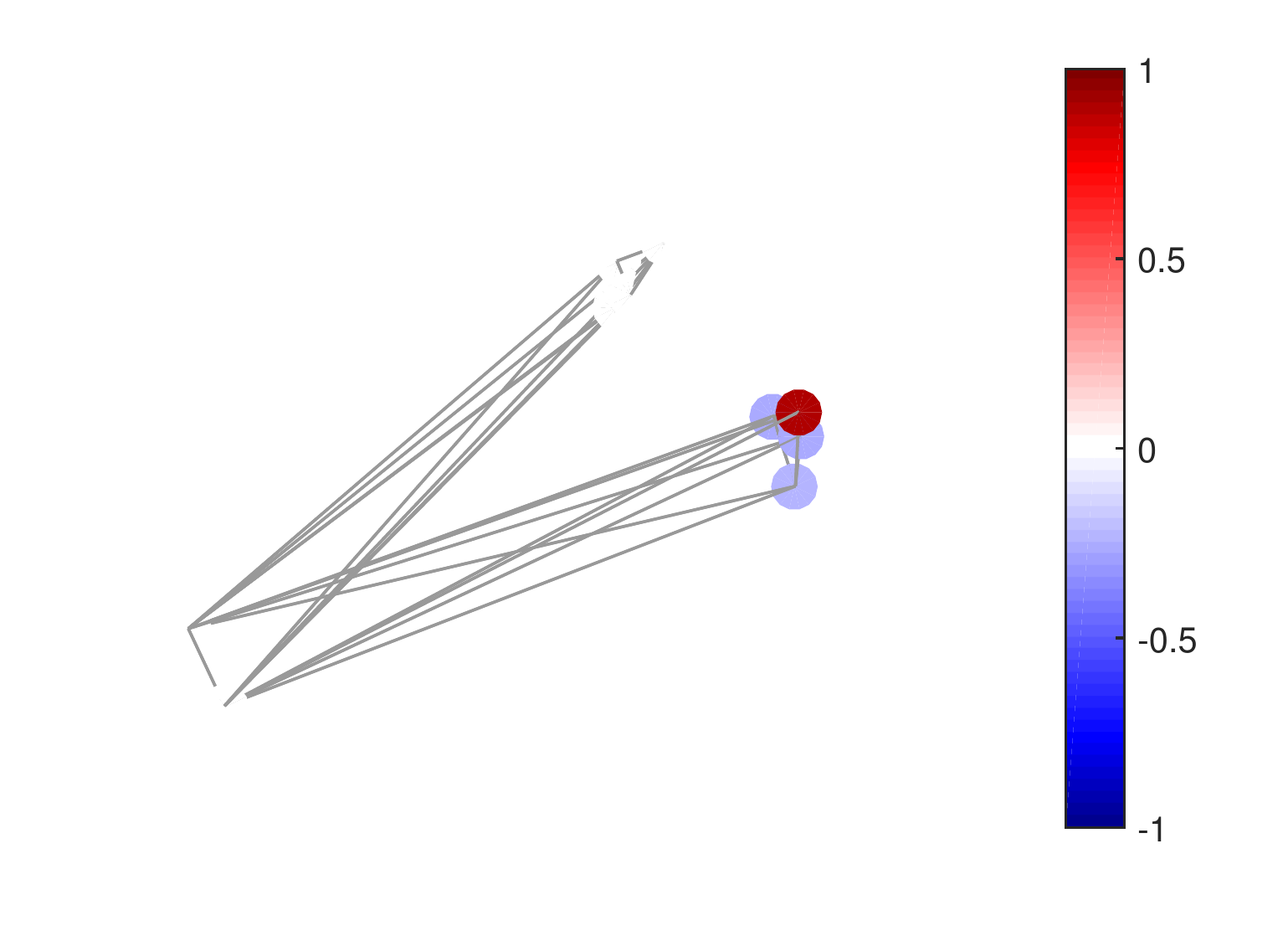}}
  \centerline{(a) PC5}\medskip
\end{minipage}
\hfill
\begin{minipage}[t]{0.49\linewidth}
  \centering
  \centerline{\includegraphics[width=8cm,height=7cm]{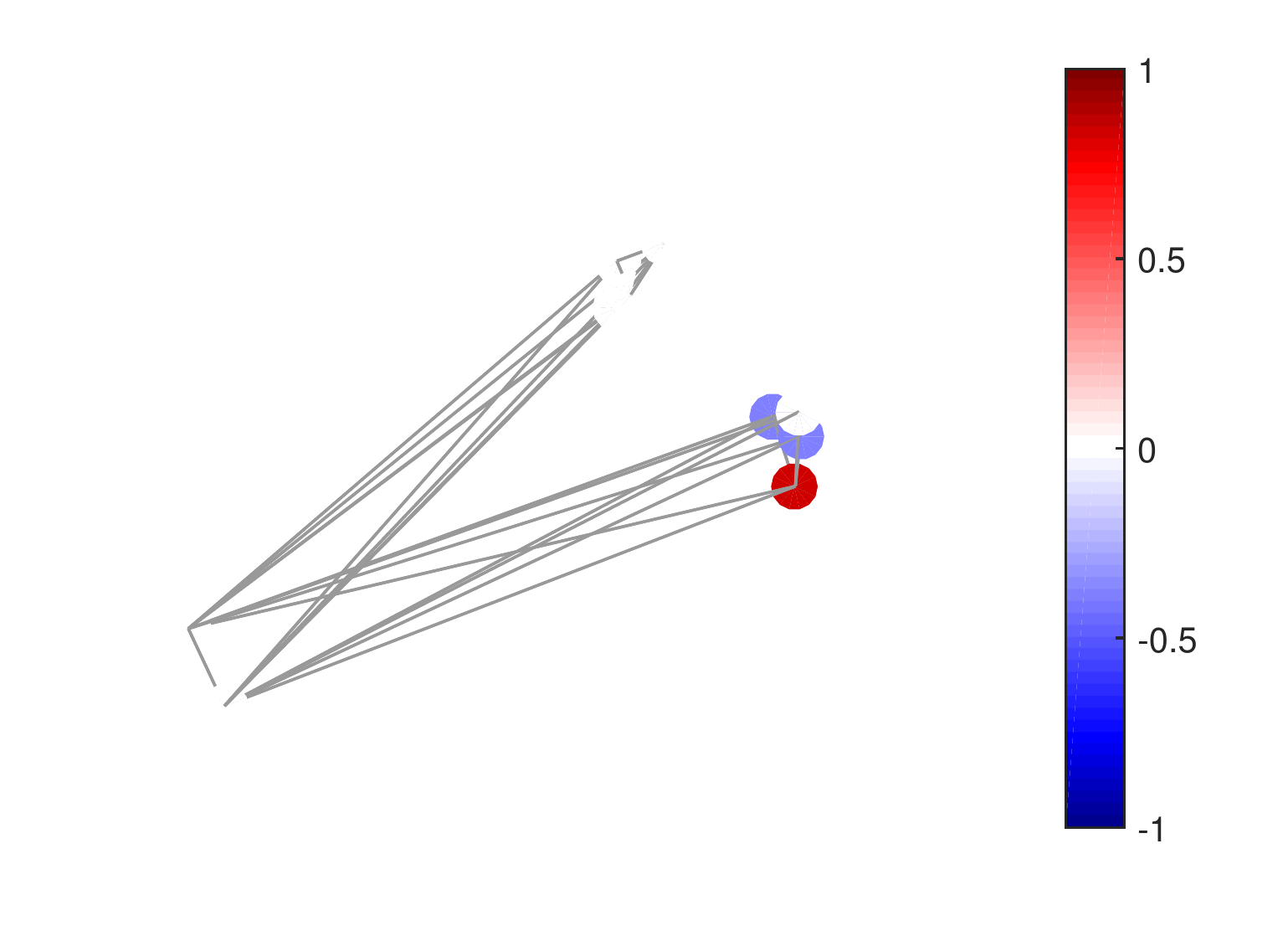}}
  \centerline{(b) PC6}\medskip
\end{minipage}
\hfill
\begin{minipage}[t]{\linewidth}
  \centering
  \centerline{\includegraphics[width=8cm,height=7cm]{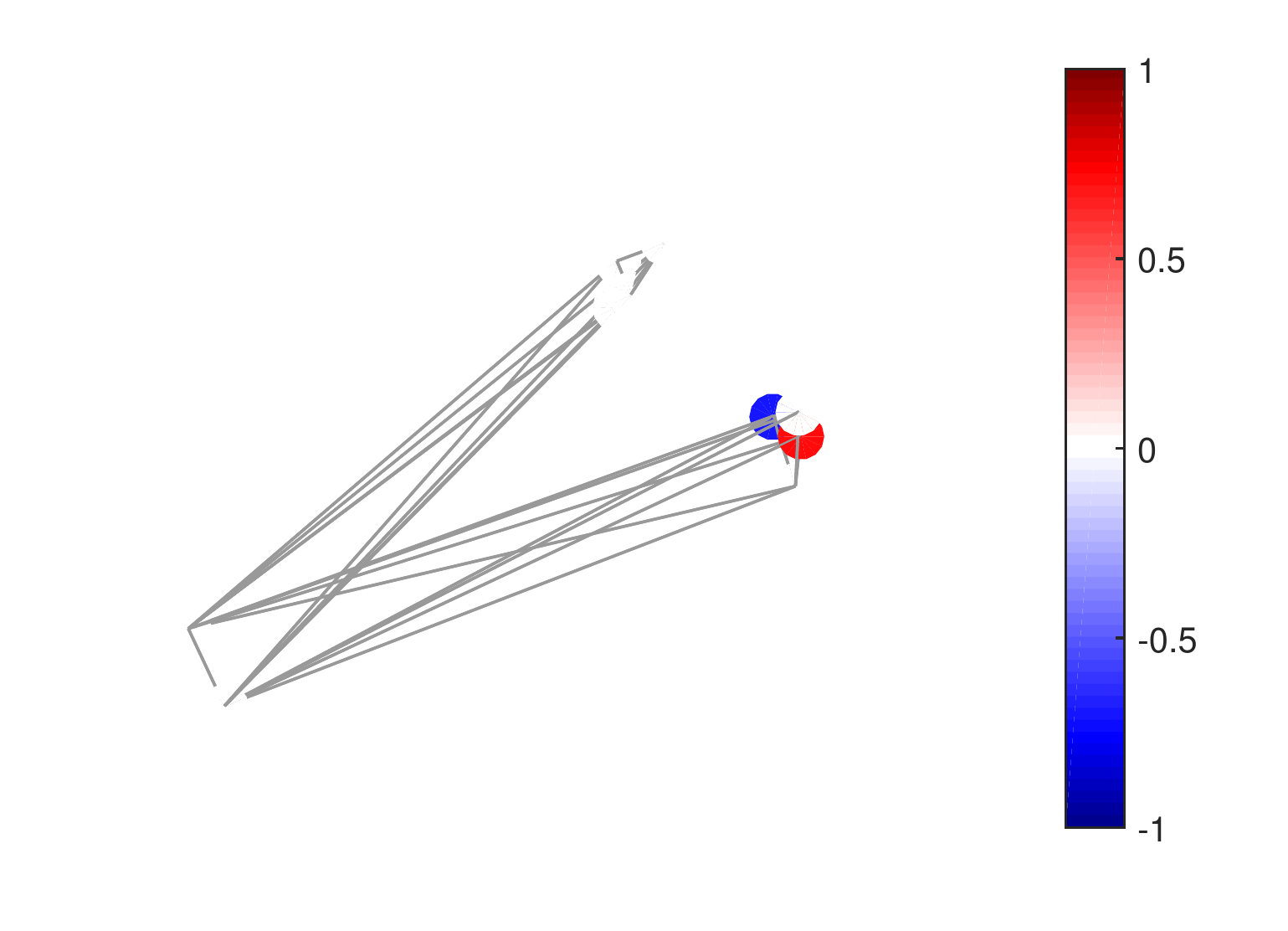}}
  \centerline{(c) PC7}\medskip
\end{minipage}
\caption{Graph visualization of analysis components of sparse GFT. The Laplacian quadratic forms are: (a) PC5, $5.7003$, (b) PC6, $5.7036$, (c) PC7, $5.7129$.}
\label{Vis}
\end{figure*}

Details and justifications of Algorithm \ref{alg:regression} can be found in~\cite{manas:sipn:2018}. In particular, the proposed Algorithm \ref{alg:regression} can obtain the same analysis components $\b_m$ as the existing approach  using eigendecomposition on the Laplacian matrix (with several details discussed in~\cite{manas:sipn:2018}). Note that one possible method of factorization of graph Laplacian is to compute the Incidence Matrix. However, for solving the optimization problem in (\ref{eq:regression1}), we do not require to compute  $\S$ explicitly. After performing some numerical transformation of (\ref{eq:regression1}), we can solve the optimization problem using FISTA, in which we do not require to compute  $\S$. See details in ~\cite{manas:sipn:2018}. 

Importantly, the advantage of Algorithm \ref{alg:regression} is that it enables incorporation of different properties into the resulting components. In particular, we consider sparse loading. This can be achieved by introducing the lasso penalty when performing the regression in Algorithm \ref{alg:regression}.  Specifically, we propose to include the following lasso penalty in (\ref{eq:regression1}):
$\sum_{m=1}^k ||\b_m||_1 $.
This leads to $\b_m$ that have sparse loading. In our proposed sparse GFT, we use this set of $\b_m$ for analysis of graph signals.
\vspace{-0.3cm}
\subsection{Analysis of sparse GFT components}
In this section, we analyze the sparse GFT components computed using Algorithm \ref{alg:regression}. In particular, we demonstrate that sparse GFT components can identify and select correlated signal sources into sub-graphs, and perform frequency analysis {\em locally} within these sub-graphs of correlated sources. Here, we provide a synthetic example to illustrate this. We construct 10 observable signal sources $X_i$ using three hidden factors. Let us assume ${V_1} \sim \mathcal{N}\left( {0,290} \right),\,\,{V_2} \sim \mathcal{N}\left( {0,300} \right)$, and ${V_3} =  - 0.01{V_1} + 0.01{V_2} + \epsilon , \,\, \epsilon  \sim \mathcal{N}\left( {0,1} \right)$, where $V_1, V_2$ and $\epsilon$ are independent. 
The observable signal sources $X_i$
are constructed as follows:
\begin{equation}
\begin{gathered}
  {X_i} = V_1+ \epsilon _i^1, \quad \epsilon _i^1 \sim \mathcal{N}\left( {0,1} \right), \quad i = 1,2,3,4 \hfill \\
  {X_i} = V_2+ \epsilon _i^2, \quad \epsilon _i^2 \sim \mathcal{N}\left( {0,1} \right), \quad i = 5,6,7,8 \hfill \\
  {X_i} = V_3+ \epsilon _i^3, \quad \epsilon _i^3 \sim \mathcal{N}\left( {0,1} \right), \quad i = 9,10 \hfill \\
  \left\{ {\epsilon _i^j} \right\}\,{\text{are independent}}, \quad j = 1,2,3, \quad i = 1,...,10 \hfill \\ 
\end{gathered}
\end{equation}
In Fig. \ref{Vis}, we depict the graph of observable signal sources $X_i$. The length of the edge ($X_i, X_j$) indicates the correlation between $X_i$ and  $X_j$: If the correlation between $X_i$ and  $X_j$ is high, the edge ($X_i, X_j$) is short.
In the figure, we depict the  5th, 6th, and 7th analysis components computed by sparse GFT. It can be observed that these components select the sub-graph containing $X_1,X_2,X_3,X_4$ which are correlated.  Furthermore, the  variations of component signals in the sub-graph are different and increase with the component number, similar to ordinary GFT except that the component signals and variations are within the sub-graph of selected sources. 
Therefore, the sparse GFT components can provide frequency analysis 
within the subgraph of correlated variables.
We also show  Laplacian quadratic forms computed for these components to quantify the variations (see Fig. \ref{Vis}).
Similar behavior is observed for 8th, 9th, 10th component which select the sub-graph containing $X_5,X_6,X_7,X_8$.

\vspace{-0.3cm}
\section{Experiments for anomaly detection}
During signal analysis, we project the signal $\x$ onto the sparse GFT components:
$\tilde{\x}[m]  = \langle \x, \b_m \rangle$. As $\b_m$ is sparse, $\tilde{\x}[m]$ is the aggregation of a few correlated signal sources, instead of all the signal sources as in ordinary GFT. This facilitates anomaly detection.
We evaluated the anomaly detection performance using the Abilene dataset. 
Details of the experiments and additional results can be found in \cite{manas:sipn:2018}. 
The comparison results shown in Table \ref{T_AUC_Abilene} suggest that sparse GFT can achieve  improved performance in an network traffic anomaly detection.
\vspace{-0.3cm}
\begin{table} [!ht]
\centering \scriptsize
\caption{Average AUC (area under curve) for anomaly detection. Details are discussed in \cite{manas:sipn:2018}.} \label{T_AUC_Abilene}
\begin{tabular}{| c | c | c | c | c | c | c | c |}
\hline
 Dataset	& PCA  	   & LPP  	 & RPCA    & RPCAG 	 & FRPCAG  & GLPCA   &  \textbf{Sparse GFT}    \\
\hline
$1^{st}$    &  $76.19$ & $65.23$ & $72.35$ & $73.51$ & $76.54$ & $71.68$ &  $\textbf{86.35}$  \\
\hline 
 $2^{nd}$   &  $64.54$ & $72.71$ & $54.26$ & $63.10$ & $65.49$ & $54.94$ &  $\textbf{82.50}$  \\
\hline
$3^{rd}$ 	&  $55.08$ & $49.55$ & $49.18$ & $54.68$ & $59.69$ & $49.20$ &  $\textbf{70.58}$ \\
\hline
\end{tabular}
\end{table}
\vspace{-0.5cm}
\section{Conclusions}
In this paper, we proposed {\em sparse GFT} which can identify and select correlated signal sources into sub-graphs, and perform frequency analysis {\em locally} within these sub-graphs of correlated sources. 
We demonstrated that sparse GFT is effective in  anomaly detection.




\end{document}